\documentclass[sigconf,10pt]{acmart}
\usepackage{xcolor}
\usepackage{dsfont}
\usepackage{algorithm}
\usepackage[noend]{algorithmic}

\AtBeginDocument{%
  }

\DeclareMathOperator*{\argmin}{arg\,min}

\setcopyright{rightsretained}
\copyrightyear{2025}
\acmYear{2025}
\acmDOI{XXXXXXX.XXXXXXX}

\acmConference[npj Wireless Technology]{npj Wireless Technology}{November}{2025}

\acmISBN{}

\begin{document}

\title{mmFlux: Crowd Flow Analytics with Commodity mmWave MIMO Radar}

\author{Anurag Pallaprolu, Winston Hurst, and Yasamin Mostofi}
\email{{apallaprolu, winstonhurst, ymostofi} @ ucsb.edu}
\affiliation{
  \institution{University of California Santa Barbara}
  \city{Santa Barbara}
  \country{USA}
}

\renewcommand{\shortauthors}{Pallaprolu, et al.}

\begin{abstract}
In this paper, we present mmFlux: a novel framework for extracting underlying crowd motion patterns and inferring crowd semantics using mmWave radar. First, our proposed signal processing pipeline combines optical flow estimation concepts from vision with novel statistical and morphological noise filtering. This approach generates high-fidelity mmWave flow fields—compact 2D vector representations of crowd motion. We then introduce a novel approach that transforms these fields into directed geometric graphs. In these graphs, edges capture dominant flow currents, vertices mark crowd splitting or merging, and flow distribution is quantified across edges. Finally, we show that 
analyzing the local Jacobian and computing the corresponding curl and divergence enables extraction of key crowd semantics
for both structured and diffused crowds. We conduct 21 experiments on crowds of up to 
20 people across 3 areas, using commodity mmWave radar. Our framework achieves high-fidelity graph reconstruction of the underlying flow structure, even for complex crowd patterns, demonstrating strong spatial alignment and precise quantitative characterization of flow split ratios. Finally, our curl and divergence analysis accurately infers key crowd semantics, e.g., abrupt turns, boundaries where flow directions shift, dispersions, and gatherings.  Overall, these findings validate mmFlux,
underscoring its potential for various crowd analytics applications.
\end{abstract}

\maketitle
\section{Introduction and Related Work}
\label{sec:intro_and_related_work}
Understanding crowd motion dynamics offers valuable insights into how a space is used. In commercial settings, analyzing customer movement can reveal which items attract attention, informing optimal product placement and store layout~\cite{greco2022ICIAP, nguyen2022ai}. In security, detecting anomalies in crowd behavior can help identify potential theft or safety threats~\cite{charankumar2024using}. Urban planners study crowd flows to design public spaces that minimize bottlenecks and ensure safe evacuations in emergencies~\cite{ALDAHLAWI2024104638}, and event organizers use similar analyses to improve venue layouts and crowd control at concerts, stadiums, and festivals~\cite{suraj2024dynamic}. More broadly, smart cities leverage crowd flow data to optimize traffic patterns, enhance public transportation, and improve overall urban mobility~\cite{kujareanpaisal2024bma}.

Historically, vision-based systems have been the primary method for extracting crowd analytics, due to their high resolution. At the same time, growing privacy concerns~\cite{guardian2025ring, lemonde2024olympic, sfchronicle2024tsa,news2024grampians, birnbaum2019privacy} have fueled interest in RF-sensing based solutions, which are naturally privacy-preserving,  with several earlier works using WiFi signals  ~\cite{shahbazian2023human, korany2021counting, depatla2015occupancy, yang2018wi, sharma2021passive}. However, the low sensing resolution of WiFi limits its applicability for crowd analytics, despite recent advances in WiFi-based sensing for other tasks ~\cite{pallaprolu2022wiffract, cai2020teaching, karanam2019tracking, miao2025wi, tan2022commodity}. On the other hand, in recent years, there has been a shift toward mmWave frequencies, driven by their central role in 5G/6G networks ~\cite{ericsson_6G_2024} as well as the increasing availability of low-cost commodity mmWave transceivers. As such, recent studies have demonstrated the potential of mmWave systems for RF sensing in general, and crowd analytics in particular~\cite{ShengleiLI2022, vaidya2024exploiting, pallaprolu2024crowd}.

\begin{figure}
    \centering
    \includegraphics[width=\linewidth]{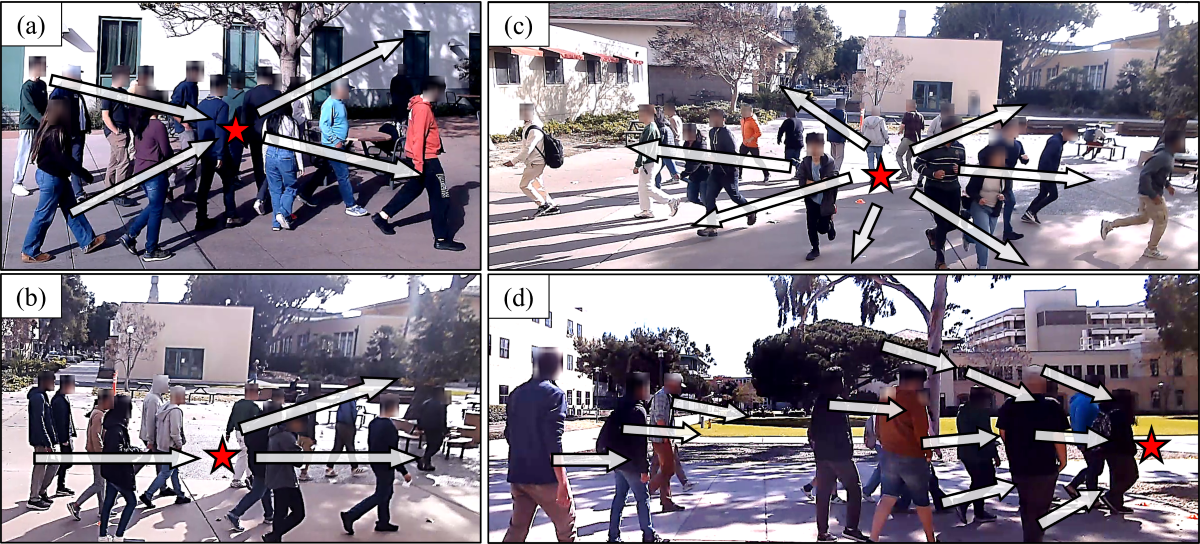}
    \caption{Sample \textit{structured crowds} showing two dominant flows (a) intersecting (b) splitting at the red star. Sample \textit{diffuse crowds} showing individuals (c) fleeing from (d) converging to red star.}
    \label{fig:sample_crowds}
\end{figure}

In this paper, we push the boundaries of mmWave-based crowd analytics by presenting mmFlux: a novel method to extract high-level motion patterns for large crowds \textbf{using a single commodity mmWave radar board}, validated through 21 experiments involving crowds of up to 20 people.
To capture a broad set of crowd behavior, we consider two categories of crowd dynamics in this paper. Specifically, we refer to crowds which have dominant flow patterns that can be well-represented by a graph as \textbf{structured crowds}. We further consider crowds that are not structured as such, but still exhibit emerging flow behaviors, which we refer to as \textbf{diffuse crowds}. 
See Fig.~\ref{fig:sample_crowds} for a few samples and Fig.~\ref{fig:different_topologies} for a comprehensive list of considered crowd dynamics.\footnote{We assume that the underlying flow structure/behavior is time-invariant, whenever there is a flow.} We next summarize our key contributions.

\noindent\textbf{Statement of Contributions:}

\noindent \textbf{a)} We introduce 
\textbf{flow fields} as a compact representation of crowd motion dynamics in mmWave crowd analytics. This approach obviates the need for end-to-end tracking of individuals in the crowd, reducing the effects of resolution limitations and occlusion. Building on traditional optical flow methods in the area of vision, our novel signal processing pipeline ensures robust flow field estimation by developing new noise-filtering and interpolation techniques tailored to the sparse and noisy nature of mmWave data.

\noindent \textbf{b)} Using the estimated flow field, we then show how to extract the high-level flow topology of structured crowds (denoted by $G(V, E)$) by constructing a directed geometric graph, $\hat{G}(\hat{V}, \hat{E})$, whose nodes represent crowd split or merge points, and edges denote dominant flow currents. Additionally, we propose a method to estimate flow split ratios at each node, quantifying the relative distribution of flows across edges. The flow topology, combined with split ratios, enables a detailed characterization of structured crowd dynamics.

\noindent \textbf{c)} For both structured and diffuse crowds, we analyze the flow field itself to extract key semantic insights. Specifically, we present a novel method that computes the curl and divergence of the flow field, using vector field analysis, enabling the identification of flow split and merge regions, sharp turns, and boundaries where dominant flow directions shift abruptly.

\noindent \textbf{d)} We validate mmFlux
through extensive testing using a commodity TI AWR2243BOOST mmWave radar board~\cite{awr2243boostug}. Specifically, we present the results of 21 real-world experiments with crowds of up to (and including) 20 people in three distinct areas that experience significant environmental noise and multipath effects. Our framework achieves high-fidelity graph reconstruction of the underlying flow structure, demonstrating strong spatial alignment even for complex crowd patterns, and further accurately estimates the flow split ratios with a Mean Absolute Error (MAE) of 0.1. We finally demonstrate how to use the curl and divergence of the flow field to partition the space into regions with different behaviors, delineate boundaries, and identify crowd semantics such as sharp turns and splits.

We next start with a comprehensive literature review.
Crowd analytics has gained significant attention in recent years, driven by the need for efficient urban planning~\cite{bendali2021recent, celes2019crowd, liao2019applying}, safety management~\cite{helbing2015saving}, and behavioral understanding in dynamic environments~\cite{adrian2020crowds}.

Vision-based methods have long dominated crowd analytics, with early work using optical flow and physics-inspired models, such as fluid mechanics and social forces, to analyze crowd dynamics~\cite{alfarano2024estimating, zhai2021optical, ali2007lagrangian, ali2008floor, mehran2010streakline, solmaz2012identifying, nayak2013vector, mehran2009abnormal, zhao2011crowd}, and later work applying deep learning methods ~\cite{shi2023flowformer++, lu2024optical, bartoli2018context, alahi2016social, farooq2022motion, habara2024floor, luo2024detecting}. Despite these advances, vision-based approaches remain vulnerable to adverse environmental conditions such as poor lighting and rain/fog, and they have provoked increasing societal resistance due to mounting privacy concerns from citizens~\cite{news2024grampians}, regulatory bodies~\cite{politico2025labor}, and advocacy groups~\cite{birnbaum2019privacy}.

As a privacy-preserving alternative to vision, researchers have explored RF-based modalities for crowd analytics. In particular, WiFi and BLE signals have been widely used for crowd monitoring~\cite{shahbazian2023human}, with prior work mainly focusing on occupancy estimation~\cite{ling2024easycount, de2022rf, korany2021counting, demrozi2021estimating, liu2019deepcount, denis2020sensing, yang2018wi,sharma2021passive, kulshrestha2019real, andion2018smart, manjappa2019estimating, depatla2015occupancy}. However, the poor sensing resolution of these frequencies greatly limits their applicability and further necessitates additional assumptions, e.g., line-of-sight crossing or use of costly and extensive transceiver networks.

On the other hand, mmWave radars have recently gained widespread attention for their privacy-preserving and weather-resilient sensing capabilities, and their high-resolution spatial sensing has been leveraged for fine-grained human motion analysis~\cite{kong2024survey}. Recent work has also started investigating crowd analytics with mmWave radar, but with most work focusing on occupancy estimation/crowd counting~\cite{pallaprolu2024crowd, hsu2023novel, li2024mmwave, ren2023grouped, hu2024mmcount, ShengleiLI2022, vaidya2024exploiting, pallaprolu2024NUCrowd}, as opposed to extracting underlying crowd flow patterns and emerging behaviors. In other words, existing mmWave-based crowd analytics studies do not ad-
dress the specific subject of this paper and their solutions are not adaptable to solve our problem of interest. For instance, several crowd counting works track individuals for counting~\cite{liu2024pmtrack, chen2024mmtai, shen2024advanced, zhu2024joint}, but they suffer from occlusion effects and resolution limits, and can thus only handle small crowd sizes~\cite{rakai2022data}. In~\cite{pallaprolu2024crowd}, we proposed a foundation for crowd counting inspired by concepts from stochastic geometry. However, the approach relied on prior knowledge of the crowd spatial usage. This underscores the need for a new framework for aggregate motion representation and analysis, in order to robustly extract underlying crowd flow patterns, which is the main motivation for this paper. 

Specifically, we start by introducing a new flow field model for mmWave crowd sensing, which shall provide a compact and robust representation of crowd dynamics, enabling subsequent crowd flow analysis.  While flow field models of crowd behavior have received significant attention in vision, such approaches remain largely unexplored for mmWave radars. A notable exception is~\cite{Zhang2024Sensors}, which focuses exclusively on anomaly detection and employs a flow
field methodology tailored to that task, thus lacking generalizable analysis and broader applicability. Beyond the domain of crowd analytics, mmWave flow field models have been developed for automotive scene flow~\cite{Ding2022RaFlow} and activity recognition for a single individual~\cite{Ding2024milliFlow}, but their task-specific training limits their applicability to crowd flow analysis.


To the best of our knowledge, no existing RF-based approach has addressed the problem of characterizing aggregate spatial crowd motion patterns. Moreover, existing methods that focus on estimating the crowd count are not applicable for characterizing the underlying crowd flow patterns.  In fact, underlying flow patterns are often treated as a prior in crowd counting. As such, the methodology proposed in this paper can also contribute to advancing the state of the art in crowd counting, opening avenues for future research.

\section{Radar Point Cloud Generation}
\label{sec:radar_pc_gen}
We start by summarizing mmWave propagation in FMCW MIMO radar systems, followed by presenting a robust methodology for generating point clouds corresponding to individuals in crowded environments. A point cloud is a collection of spatial data points representing reflective surfaces in the sensing field of view, and it serves as a key intermediate representation in several radar-based sensing works. Traditional approaches for point cloud generation rely on range-angle 2D-FFT, combined with Constant False Alarm Rate (CFAR) detection~\cite{liu2024pmtrack, canil2023oracle}. However, they require careful parameter tuning and are sensitive to environmental variations. To improve robustness, we instead propose a modified approach in this section by performing multi-scale range-only peak detection over time before applying targeted Angle-of-Arrival (AoA) estimation at identified sensing depths. This method can reduce false detections while providing precise spatial representations of crowd dynamics. We next begin with a summary of FMCW radar transmission, followed by presenting the details of our approach to point cloud generation. 

Consider a mmWave FMCW MIMO radar system consisting of \(N_{\text{TX}}\) transmitters (TX) and \(N_{\text{RX}}\) receivers (RX), where each TX periodically emits a chirp sinusoid with a frequency that increases linearly from \( f_0 \). The transmitted frequency is given by $f_{\text{TX}}(t) = f_0 + St,\ 0 \leq t < T_{c}$, where \( S = B/T_c \) denotes the chirp slope, \( B \) is the bandwidth, and \( T_c \) is the chirp duration. The complex baseband signal of the \( m^{\text{th}} \) chirp of the \( i^{\text{th}} \) TX is then given by $s_{\text{TX}}^{(i)}(m,t) = \sqrt{P_{\text{TX}}} \exp \big(j\{2\pi f_0 t + \pi S t^2\}\big)$, where \( P_{\text{TX}} \) is the transmit power.

Next, consider an object located at a distance $d_j[m]$ from the \( j^{\text{th}} \) RX. The signal received at the \( j^{\text{th}} \) RX after scattering is given by $s_{\text{RX}}^{(i, j)}(m,t) = s_{\text{TX}}^{(i)}(m,t - \tau_{j})/d_j[m]^2$, where \( \tau_{j} = 2d_j[m]/c \) is the round-trip time of flight, and $c$ is the speed of light. The received signal is then mixed with the transmitted chirp\footnote{We use a TDM-MIMO setup where TXs emit chirps sequentially, avoiding the need to resolve signal arrivals from different TXs.} at each RX to yield the intermediate frequency (IF) signal ~\cite{stove1992linear, rao2017introduction}: $s_{\text{IF}}^{(i, j)}(m,t) = s_{\text{RX}}^{(i, j)*}(m,t) \times s_{\text{TX}}^{(i)}(m,t)$.

To improve angular resolution, we construct a virtual array by pairing each TX-RX element in order to synthesize an extended rectangular aperture, as is commonly done in prior work~\cite{kong2022m3track}. Defining virtual indices $(p, q)$ with $p \in \{1, \dots, N_x\}$ and $q \in \{1, \dots, N_y\}$, we map each TX-RX pair $(i, j)$ to a virtual array element $(p, q)$ via a convolution of the TX array pattern over the RX array~\cite{TI_SWRA554A}. Reordering the received data accordingly forms a planar array, effectively increasing spatial samples from $N_{\text{RX}}$ to $N_xN_y = N_{\text{TX}}N_{\text{RX}}$ and refining angular resolution beyond a single TX-RX pair.

Let \( s_{\text{IF}}^{(p,q)}[m, n] \) denote the discrete IF signal, where $n$ is the Analog-to-Digital Converter (ADC) bin index, \( N_r = T_c/T_r \) represents the number of discrete range bins with \( T_r \) denoting the ADC sampling duration, $N_c$ denotes the chirp duration, and \( T_f = N_cT_c \) denotes the total time duration. Traditional point cloud generation methods~\cite{pegoraro2021real, canil2023oracle, yan2023mmgesture} employ a range-Doppler-angle FFT followed by CFAR thresholding in order to extract point clouds over time. However, these methods exhibit several limitations. CFAR assumes a predefined radar noise distribution and requires the inversion of complex nonlinear operations to determine adaptive threshold values~\cite{rohling1983radar,kronauge2013fast}. Moreover, noise characteristics are environment-dependent, necessitating recalibration for each deployment scenario~\cite{hong2014performance}. Lastly, CFAR applies a fixed-size sliding window~\cite{rohling2011ordered} for threshold estimation, lacking a multi-scale perspective. This limitation reduces its effectiveness in detecting targets with varying spatial extents or in cases where noise characteristics change across scales.

In this paper, we then adopt a modified approach that enhances the quality of point cloud generation by first performing range detection and then applying targeted AoA estimation only at the detected ranges, thereby minimizing the false positives typically encountered in direct detection over 2D range-angle spectra. Specifically, we leverage the pipeline described in~\cite{pallaprolu2024crowd} to first generate Binary Trace Maps, which encode the occupancy of visible agents across range and time. This method involves adaptive multi-scale range peak detection that automatically adjusts to varying target dimensions and spatial characteristics, enabling more effective denoising across different spatial scales. For completeness, we next outline the prerequisite steps.

First, we compute the Fourier transform along the ADC axis, \textit{i.e.,} the range-FFT:\footnote{Note that range localization of visible agents does not utilize the virtual array, and thus we drop the $(p, q)$ indices on $s_{\text{IF}}[m,n]$.}
\begin{align}
    R_{\text{IF}}[m,r] &= \sum\nolimits_{n=0}^{N_r-1} s_{\text{IF}}[m,n] \exp\Bigl(-j\frac{2\pi n r}{N_r}\Bigr).
\end{align}

Next, instead of the traditional Doppler Fourier transform step, we compute the Wrapped Phase Spectrum:
\begin{align}
\label{eq:wrapped_phase_transform}
    \zeta_{\text{IF}}[z, w, r] &= \sum\nolimits_{i=w}^{w+W} \text{Arg}(R_{\text{IF}}[i, r]) \exp\Bigl(-j\frac{2\pi z i}{W}\Bigr),
\end{align}

\noindent where $\text{Arg}(Ae^{j\gamma}) = \gamma\ \text{mod}\ 2\pi$ denotes the wrapped phase of a complex number, and $w$ represents the index of the starting chirp in a sliding window of $W$ chirps. We then compute the spectral bandwidth along the $z$-dimension of $\zeta_{\text{IF}}[z, w, r]$ as:
\begin{align}
    \mathcal{H}[w, r] = \Gamma_z(\zeta_{\text{IF}}[z, w, r]),
\end{align}

\noindent where $\Gamma_z(f(z)) \in \mathbb{R}$ is a suitable measure of the bandwidth of $f(z)$. Subsequently, we binarize $\mathcal{H}$ using an ensemble of pretrained Gradient Boosted Machines~\cite{pallaprolu2024crowd} to obtain the Binary Trace Map $\mathcal{H}_{\text{b, vis}}[w, r]$, resulting in a set of detections:
\begin{equation}
\label{eq:GBM_range_detections}
    \mathcal{D}_R[w] = \{r_i \mid \mathcal{H}_{\text{b, vis}}[w, r_i] = 1,\ 1 \leq i  \leq K[w]\},
\end{equation}
where $K[w] = \sum_{r=0}^{N_r}\mathcal{H}_{\text{b, vis}}[w, r]$ represents the number of agents visible to the radar at chirp window $w$.

Having established the range detection approach, which localizes activity only along the range dimension over time, we next extend it by reintroducing the virtual array to perform AoA estimation only at the sensing depths detected in $\mathcal{D}_R[w]$ for each $w$, thereby robustly constructing point clouds of crowd activity over time. First, we denote the range-Doppler transform for the array element $(p, q)$ over chirp window $w$ as
\begin{align*}
  Q_{\text{IF}}^{(p, q)}[w, v, r] = \mathcal{F}_{m, n} \big\{ s_{\text{IF}}^{(p,q)}[m, n] \mathds{1}_{w \leq m < w+W} \big\}.  
\end{align*}
 
For each range detection in Eq.~(\ref{eq:GBM_range_detections}), we then perform a 2D angle-FFT along the aperture dimensions $(p, q)$. Since Doppler information is not utilized in this work, we average over the Doppler axis. Furthermore, given that off-the-shelf radar boards inherently have poor elevation resolution, we sum over the elevation dimension after the angle-FFT step to obtain the effective range-azimuthal angle spectrum:
\begin{align}
\label{eq:azim_spectrum}
    \mathcal{A}[w, \theta, r] &= \sum\nolimits_{\phi} A_{\text{IF}}[w, \theta, \phi, r], \text{ where}\nonumber \\
    A_{\text{IF}}[w, \theta, \phi, r] &= \frac{1}{N_c} \sum\nolimits_{v=0}^{N_c - 1} \mathcal{F}_{p, q} \{ Q_{\text{IF}}^{(p,q)}[w, v, r] \}.
\end{align}

Finally, we construct our point cloud by iterating over each detected range $r_i \in \mathcal{D}_R[w]$ to calculate the corresponding bearing angle, $\theta_i$, that maximizes the azimuthal spectrum $\mathcal{A}$ of Eq.~(\ref{eq:azim_spectrum}) at range $r_i$:
\begin{equation}
\label{eq:pc_detections}
    \mathcal{D}[w]\!=\!\{(r_i, \theta_i=\arg\max_{\theta}\mathcal{A}[w, \theta, r_i])
    \!\mid\!r_i\!\in\!\mathcal{D}_R[w]\},
\end{equation}

\noindent which results in the final point cloud representation. While this representation does not fully resolve individual trajectories due to occlusions, noise, and overlapping detections, it effectively captures the overall crowd dynamics, enabling the inference of collective movement trends, as we shall see. We next introduce the concept of a flow field, which helps transform these discrete, sporadic, and noisy observations into efficient spatial representations of crowd motion patterns, enabling accurate reconstruction of crowd flow topologies and inference of underlying flow behaviors and semantics.

\begin{figure*}
    \centering
    \includegraphics[width=\linewidth]{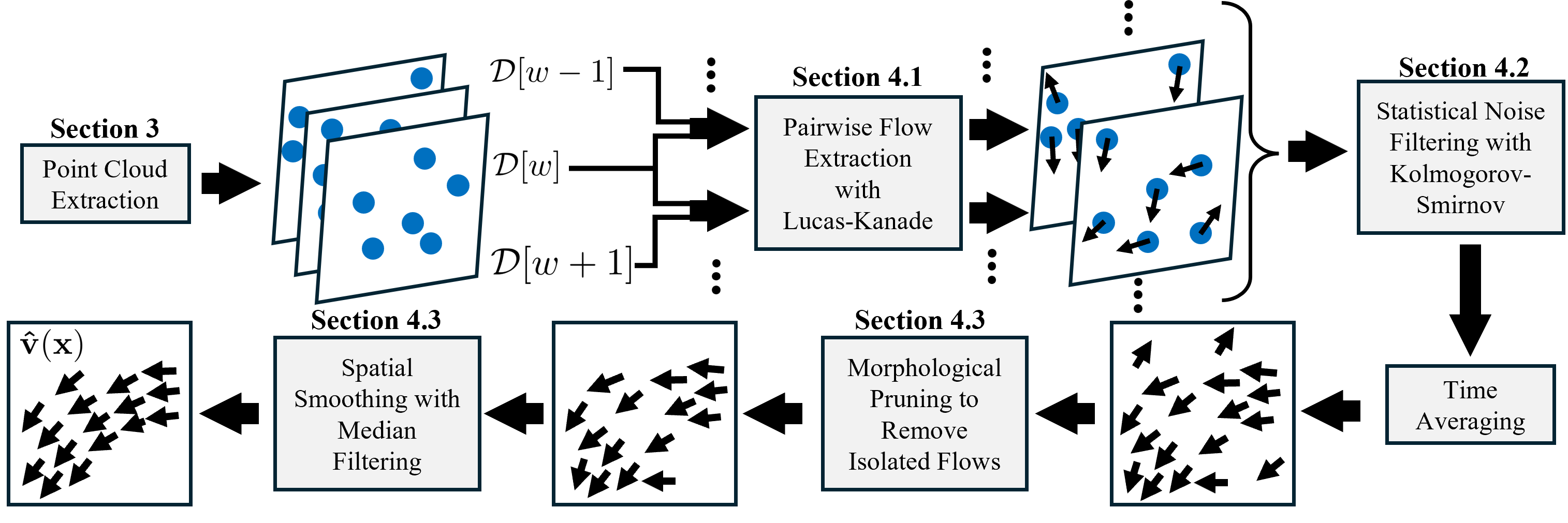}
    \caption{Illustration of our proposed flow field generation pipeline. We extract flows between consecutive point clouds and filter locations with inconsistent flows over time using a Kolmogorov-Smirnov test. We then compute the average flow, apply morphological pruning to remove isolated flow regions, and perform local spatial smoothing to obtain the final estimated flow field, which serves as the starting point for the analysis presented in Sec.~\ref{sec:crowd_flow_analysis}.}
    \label{fig:pipeline}
\end{figure*}

\section{Flow Field Estimation}
\label{sec:ff_synthesis}
The core step between point cloud generation and graph reconstruction is the estimation of the flow field, a 2D vector field, $\mathbf{v}(\mathbf{x}): \mathbb{R}^2 \to \mathbb{R}^2$, which characterizes the aggregate movement of individuals in the crowd. In this section, we describe our method for estimating $\mathbf{v}(\mathbf{x})$. We first leverage a classical optical flow method to extract instantaneous flows at each time step from the sequence of point clouds (Sec.~\ref{sec:radar_pc_gen}). We then present novel filtering and interpolation techniques tailored for the sparse, noisy nature of mmWave data to construct an estimated flow field, $\mathbf{\hat{v}}(\mathbf{x})$, which is then used in Sec.~\ref{sec:crowd_flow_analysis} as part of our proposed methodology to uncover crowd flow topologies and extract crowd flow semantics. Fig.~\ref{fig:pipeline} outlines our approach.

\subsection{Pairwise Flow Estimation}
\label{subsec:pwf_generation}
As illustrated in Fig.~\ref{fig:pipeline}, the first step towards estimating the flow field, $\mathbf{v}(\mathbf{x)}$, is to transform the sequence of point clouds into a sequence of sparse vector fields. More specifically, for each point $\mathbf{x}_{w,i} =(r_i[w] \text{cos}(\theta_i[w]), r_i[w]\text{sin}(\theta_i[w]))$ in a point cloud $\mathcal{D}[w]$, we estimate the flow at that point, $\hat{\mathbf{v}}_{w,i}$, by comparison to the subsequent point cloud, $\mathcal{D}[w+1]$. If $\mathbf{x}_{w,i}\in \mathcal{D}[w]$ evolves according to a flow field $\mathbf{v}(\mathbf{x})$, then at the next radar sampling window, $w+1$, we will find a radar detection at a location $\mathbf{x}_{w+1, i'} = \mathbf{x}_{w,i} + \mathbf{v}(\mathbf{x}_{w,i})$, wherein we assume that time in the flow field is normalized with respect to the radar sampling rate. Thus, if there were no errors in observation and if point association from one point cloud to the next were known, we could precisely recover the flow field at the observed point as $\mathbf{v}(\mathbf{x}_{w,i}) = \mathbf{x}_{w+1,i'} - \mathbf{x}_{w,i}$.

However, associating points between point clouds is challenging ~\cite{pearce2023multi, rakai2022data} due to the large number of matches required, compounded by missed detections and false alarms. Furthermore, the presence of noise implies that even if associations are known, the resulting flow estimation would be inexact.

To address these challenges, we leverage well-established optical flow methods from computer vision, which are designed to infer motion between successive image frames. Specifically, we employ the classical Lucas-Kanade algorithm~\cite{lucas1981iterative, opencv_lk} on successive point clouds, \(\mathcal{D}[w]\) and \(\mathcal{D}[w+1]\) to find the pairwise flow (PWF) $\mathbf{\hat{v}}_{w,i}$ for all $\mathbf{x}_{w,i}\in \mathcal{D}[w]$. As Lucas-Kanade requires inputs in image format, we first discretize the space and create a binary occupancy map, $\mathcal{H}[w](\mathbf{x})\in \{0,1\}$, where $\mathcal{H}[w](\mathbf{x}) = 1$ if there exists a point in \(\mathcal{D}[w]\) at location $\mathbf{x}$. For each point $\mathbf{x}_{w,i}\in \mathcal{D}[w]$, the Lucas-Kanade algorithm then estimates the corresponding flow, $\mathbf{\hat{v}}_{w,i}$, by solving the following least-squares optimization problem:
\begin{align}
\label{eq:pwf_gen}
    \mathbf{\hat{v}_{w,i}} = \argmin_{\mathbf{h}} \hspace{-0.15in}\sum_{\mathbf{y} \in \mathcal{N}_o(\mathbf{\mathbf{x}_{w,i}})}\hspace{-0.1in} \left\lVert \mathcal{H}[w+1](\mathbf{y} + \mathbf{h}) - \mathcal{H}[w](\mathbf{y}) \right\rVert^2.
\end{align}

\noindent Here, \(\mathcal{N}_o(\mathbf{\mathbf{x}_{w,i}})\) denotes a local neighborhood around \(\mathbf{\mathbf{x}_{w,i}}\), and $\mathbf{\hat{v}_{w,i}}\in \mathbb{R}^2$ is the displacement vector that minimizes the intensity difference between successive frames. 

Due to the sparsity of point clouds, each point in space will not have an estimated flow associated with it at every time step, but after computing the PWFs for all consecutive pairs of point clouds, we can define a (possibly empty) set of PWFs at each point, $\psi(\mathbf{x})= \{\mathbf{\hat{v}}_{w,i} | \mathbf{x}_{w,i} = \mathbf{x}\}$, which gives all estimated flows at $\mathbf{x}$ over time. We next propose a method of statistical analysis on these sets to discard unreliable flows.

\subsection{Flow Field Denoising with Kolmogorov-Smirnov Statistical Tests}
\label{subsec:ks_denoising}
To produce the best estimate of the flow field at a point, $\mathbf{x}$, we must incorporate information from all relevant PWFs contained in the set $\psi(\mathbf{x})$. A straightforward approach is to directly estimate $\mathbf{v}(\mathbf{x})$ by averaging over $\psi(\mathbf{x})$:
\begin{align}\label{eq:TAF}
\mathbf{v}_{\text{TAF}}(\mathbf{x}) = \begin{cases}
    \frac{1}{|\psi(\mathbf{x})|} \sum_{\mathbf{\hat{v}}_{w,i}\in \psi(\mathbf{x})} \mathbf{\hat{v}}_{w,i} & \text{If } |\psi(\mathbf{x})| >0\\
    0 & \text{Otherwise}
\end{cases},
\end{align}

\noindent and we refer to this initial estimate as the Time-Averaged Flow Field (TAF). However, this direct approach is ineffective as it fails to distinguish between valid flows generated by crowd movement and spurious returns due to environmental factors~\cite{zavorka2024characterizing, cardamis2025leafeon} or multipath effects.

To address these challenges, we propose a novel denoising strategy that allows us to generate high-integrity flow fields. Our key insight is that a truly noisy flow exhibits no consistent directional preference, resulting in flow angles that are uniformly distributed across all possible directions. In contrast, legitimate flow patterns demonstrate consistent directionality, even in the presence of measurement noise.

We formalize this analysis by modeling each PWF in $\psi(\mathbf{x})$ as the sum of a true crowd flow component plus a noise term: $\mathbf{\hat{v}}_{w,i} = \mathbf{v}(\mathbf{x}) + \mathbf{\eta}$. Here, $\mathbf{\eta}$ is the realization of a zero-mean isotropic 2D Gaussian random vector with covariance $\sigma^2(\mathbf{x})I_2$, where $I_2$ is the $2\times 2$ identity matrix, and the noise power at $\mathbf{x}$, $\sigma^2(\mathbf{x})$, is determined by spatially varying environmental factors. If the noise power is much greater than the magnitude of the true flow, $\mathbf{v}(\mathbf{x})$, then the PWFs in $\psi(\mathbf{x})$ are approximately drawn from a zero-mean 2D Gaussian random variable, and consequently, the empirical distribution of the angles of the flows is consistent with a uniform distribution over $[0, 2\pi)$~\cite{mardia2009directional}. When this is the case, the PWFs at $\mathbf{x}$ are discarded since they do not carry reliable information about the true flow field, $\mathbf{v}(\mathbf{x)}$.

To this end, we employ a Kolmogorov-Smirnov (KS) statistical hypothesis test~\cite{massey1951kolmogorov}. At each spatial location $\mathbf{x}$, we calculate the empirical CDF of the PWF angles:
\begin{align}
    \Omega_{\mathbf{x}}(\alpha) = \frac{1}{|\psi(\mathbf{x})|} \sum\nolimits_{\mathbf{\hat{v}}_{w,i}\in \psi(\mathbf{x})} \mathds{1}_{\angle \hat{\mathbf{v}}_{w,i} \leq \alpha },
\end{align}

\noindent where $\angle \mathbf{\hat{v}}_{w,i}$ denotes the orientation of the flow vector. We then test whether the empirical CDF of flow angles, $\Omega_{\mathbf{x}}(\alpha)$, differs significantly from the CDF of a uniform distribution on $[0, 2\pi)$, denoted by $F_{\text{unif}}(\alpha)$. The KS test computes the test statistic, $D_{\mathbf{\mathbf{x}}} = \sup_{\alpha \in [0, 2\pi)} |\Omega_{\mathbf{x}}(\alpha) - F_{\text{unif}}(\alpha)|,$ and then uses the complementary CDF of the Kolmogorov distribution, $\kappa$, to compute the $p$-value, $p_{\mathbf{x}} = \kappa(D_\mathbf{x})$~\cite{marsaglia2003evaluating}.

We next establish a threshold, $p_{\text{th}}$, for classification: points with $p_{\mathbf{x}} > p_{\text{th}}$ are classified as having noisy flow, as we cannot reject the hypothesis that their flow directions follow a uniform distribution. Conversely, points with $p_{\mathbf{x}} \leq p_{\text{th}}$ exhibit statistically significant directional preference, indicating meaningful flow. 
By applying this test to each point in our field of view, we generate a binary mask defined as $M(\mathbf{x}) = \mathds{1}_{p_{\mathbf{x}} \leq p_{\text{th}}}$. We use this mask to filter out noisy points from the TAF, resulting in a Denoised Flow Field (DFF), $\mathbf{v}_{\text{DFF}}(\mathbf{x}) = M(\mathbf{x}) \odot \mathbf{v}_{\text{TAF}}(\mathbf{x})$, where $\odot$ indicates the Hadamard product. The threshold, $p_{\text{th}}$, represents a tunable parameter that balances between retaining meaningful flow information and eliminating noise. 

Overall, our proposed KS filtering significantly reduces noise by testing for the temporal consistency of flows at each point. To further refine our flow field estimate, we next introduce complementary filtering steps based on the spatial correlations naturally present in crowd movement.

\subsection{Flow Pruning and Smoothing}
\label{subsec:flow_smoothing}
As a final step in the flow field estimation process, we exploit the spatial correlation present in natural crowd flows to refine $\mathbf{v}_{\text{DFF}}(\mathbf{x})$. In particular, points along dominant movement paths form contiguous regions, so that points with valid flows which are possibly erroneously rejected as noise are often embedded in regions that, overall, show strong, well-defined flows. To account for this spatial correlation, we perform morphological pruning and spatial smoothing.

We employ morphological pruning to account for small clusters of spurious flows that remain detached from the dominant flow topology. These artifacts typically arise due to transient reflections, multipath effects, or local inconsistencies in radar detections, and they do not meaningfully contribute to the flow structure. To address this, we employ a connected-component analysis~\cite{stockman2001computer} to filter out small, disjoint flow regions. Given a predefined size threshold of $\mu_S \text{ m}^2$, this operation removes any contiguous clusters of flow vectors that fall below $\mu_S$, preserving only the large-scale flow structures. Fig.~\ref{fig:pipeline} illustrates the impact of this procedure.

As a final filtering step, we eliminate erratic flow variations that are atypical in real crowds through median filtering, a widely recognized technique for improving optical flow estimation~\cite{sun2010secrets}. We apply two consecutive rounds of median filtering with a window size of  $\mu_M \text{ m}^2$. Specifically, in the first round, we compute the median considering only neighboring non-zero flow values, while in the second, we apply standard median filtering. Beyond smoothing the flow, this two-step process has an overall effect similar to morphological closure~\cite{soille1999morphological}, a classical image processing technique for noise removal that combines dilation and erosion. The first median filter acts as a dilation, interpolating flow at zero-flow points with at least one non-zero neighbor. The second functions similarly to erosion, setting the flow to zero if more than half of the neighboring flows are zero.

By combining statistical KS-based filtering, morphological pruning, and spatial smoothing, we obtain our final estimated flow field, $\mathbf{\hat{v}}(\mathbf{x})$, that is significantly more robust to spurious noise while retaining meaningful motion patterns (see Fig.~\ref{fig:denoising_perf} and related discussion in Sec.~\ref{sec:discussion}). This refined flow field provides the basis for constructing a compact graph representation of crowd movement, capturing dominant flow pathways and connectivity patterns, which we explore next.

\begin{figure*}
    \centering
    \includegraphics[width=\linewidth]{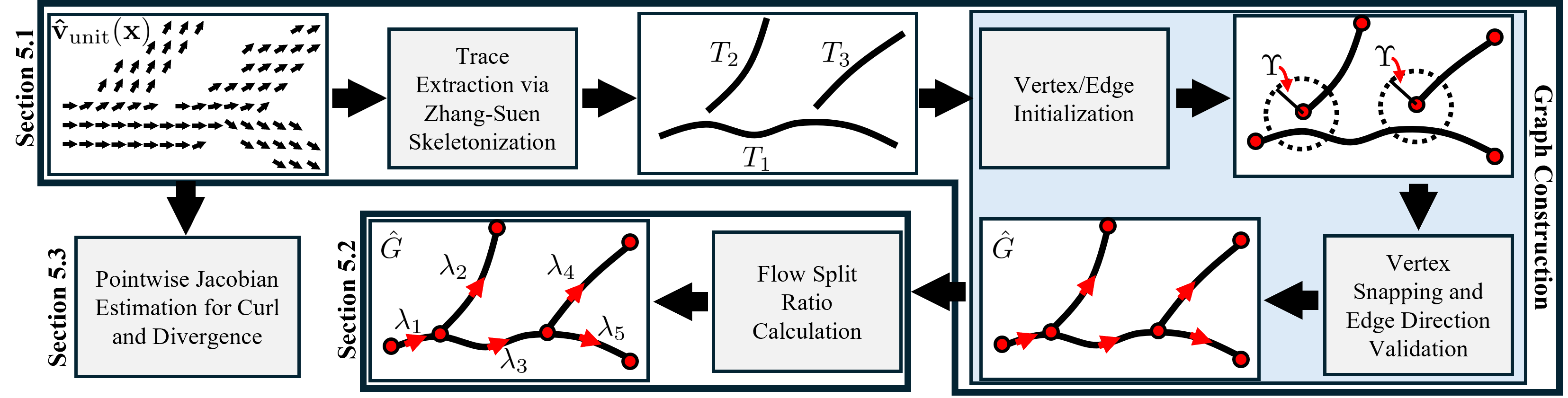}
    \caption{Illustration of our proposed robust graph reconstruction pipeline. To complement the extracted directed geometric graph, $\hat{G}$, we find edge weights, $\lambda_e$, as well as semantic details captured by curl and divergence fields.}
    \label{fig:graph_const_pipeline}
\end{figure*}

\section{Crowd Flow Analysis}
\label{sec:crowd_flow_analysis}
In this section, we introduce our proposed methodology for extracting the crowd flow topology, $\hat{G}(\hat{V}, \hat{E})$ from  the estimated flow field, $\mathbf{\hat{v}}(\mathbf{x})$, as illustrated in Fig.~\ref{fig:graph_const_pipeline}. We first present our topology reconstruction pipeline, specifically designed for structured flows, followed by a robust estimation framework for flow split ratios at the detected vertices, $\hat{V}$. We then derive the divergence and curl fields of $\mathbf{\hat{v}}(\mathbf{x})$, which provide complementary semantic insights into accumulation, dispersion, and rotation in the underlying flow.

\subsection{Flow Topology Reconstruction}
\label{subsec:flow_skeleton}
We infer the flow topology from the final flow field by employing a systematic approach that extracts the core structure and identifies critical nodes and edges, 
as shown in Fig.~\ref{fig:graph_const_pipeline}. We begin by refining our final flow field, $\mathbf{\hat{v}}(\mathbf{x})$, through a normalization process. Specifically, for each point $\mathbf{x}$, we normalize the flow vector to unit magnitude: $\mathbf{\hat{v}}_{\text{unit}}(\mathbf{x}) = \mathbf{\hat{v}}(\mathbf{x})/||\mathbf{\hat{v}}(\mathbf{x})||$ if $\mathbf{\hat{v}}(\mathbf{x}) \neq \mathbf{0}$, else $\mathbf{\hat{v}}_{\text{unit}}(\mathbf{x}) = \mathbf{0}$. We then consider the scalar field $s(\mathbf{x}) = ||\mathbf{\hat{v}}_{\text{unit}}(\mathbf{x})||$, which effectively represents a binary image where the non-zero regions correspond to the presence of flow. To extract the dominant flow structure, we use the Zhang-Suen skeletonization algorithm~\cite{zhang1984fast, chen2012improved} on $s(\mathbf{x})$, which iteratively applies local connectivity rules to erode boundary pixels while preserving the topology (Fig.~\ref{fig:graph_const_pipeline}). This process yields a flow skeleton $s_{\text{skel}}(\mathbf{x})$, which is a single-pixel wide binary representation that captures the essential flow pathways. We next implement a path tracing algorithm that traverses the skeleton, extracting these pathways which then serve as precursors to the edges in our flow graph.

To this end, we utilize an 8-connected neighborhood convolution kernel~\cite{godfried2008grids} to examine the local pixel neighborhood, enabling the systematic identification of critical points on the skeleton. Specifically, we classify branch points as pixels with more than two neighbors, and start/end points as pixels with exactly one neighbor. The algorithm then constructs skeletal traces, defined as connected paths within the skeleton, and represented as an ordered sequence of coordinates \(T_i = \{\mathbf{t}_1, \mathbf{t}_2, \dots \mathbf{t}_m\}\). We employ a systematic traversal methodology by iteratively exploring neighboring pixels, starting from each start/end point and subsequently from any unvisited branch points. Thus, $\mathbf{t}_1$ and $\mathbf{t}_m$ are either branch points or start/end points, whereas any other $\mathbf{t}_j\in T_i$ has exactly two neighbors. A global visited coordinate set is maintained throughout to ensure that each pixel is processed only once, preventing redundant traversal. We denote the set of all such skeletal traces as \(\mathcal{T} = \{T_1, T_2, \dots, T_n\}\), forming the basis for building our directed geometric graph.

Constructing a directed geometric graph from skeletal traces presents two significant challenges: establishing connectivity at flow junctions and ensuring that the edge directionality aligns with the flow field. We achieve this through a structured approach for graph construction, as detailed in Fig.~\ref{fig:graph_const_pipeline}. First, the vertex set, $\hat{V}$, is initialized using the endpoints of skeletal traces in $\mathcal{T}$, and each trace $T$ is added as an edge, $e = (\mathbf{t}_s, \mathbf{t}_e, T)$, where $\mathbf{t}_s$ and $\mathbf{t}_e$ denote the start and end points of the trace. Next, a vertex-trace proximity analysis is performed, where each vertex is examined for close proximity to existing traces; if a vertex lies within a distance threshold, $\Upsilon$, the corresponding trace is partitioned, and new edges are created to incorporate the vertex. Finally, a trace direction validation step ensures consistency with the underlying flow field by reversing the orientation of edges misaligned with the flow direction. The resulting graph, $\hat{G} = (\hat{V}, \hat{E})$, encodes the topological structure of the skeletal traces while integrat-
ing spatial and directional constraints. We note that points on edges in the geometric graph are naturally assigned a direction by referencing $\mathbf{\hat{v}}_{\text{unit}}(\mathbf{x})$.

We next show the use of the reconstructed geometric graph, $\hat{G}$, in conjunction with the point clouds, $\mathcal{D}$ (see Eq.~(\ref{eq:pc_detections})), to determine the flow split ratios at each vertex in $\hat{V}$.

\subsection{Flow Split Ratio Estimation} 
\label{subsec:flow_rate_split_estimation}
In this subsection, we augment our estimated crowd flow topology, $\hat{G}$, by finding flow split ratios, $\lambda_e \in [0,1]$, for each edge $e \in \hat{E}$. These ratios represent the relative flow distribution along the edges of $\hat{G}$. To achieve this, we employ poly-line buffering~\cite{shapely_buffer} with a width parameter $\rho$ to generate a bounding polygon, $\mathcal{B}_e$, around the skeletal trace $T$ corresponding to the edge $e$. To find the relative occupancy of each edge, we then determine the number of points over all point clouds, $\mathcal{D}[w]$, that lie within the bounding polygon, and we divide this number by the total area of the polygon to get the number of returns per unit area associated with each edge. Formally, we find unnormalized weights $\tilde{\lambda}_e = \sum_{\mathbf{x}_{w,i}\in \mathcal{D}} \mathds1_{\mathbf{x}_{w,i} \in \mathcal{B}_e}/|\mathcal{B}_e|$, where $\mathcal{D} = \cup_{w=0}^{W} \mathcal{D}[w]$ and $|\mathcal{B}_{e}|$ denotes the area of the bounding polygon $\mathcal{B}_e$. Finally, for each vertex $q\in \hat{V}$, we identify the set of outgoing edges, $\hat{V}_{\text{out}}(q)$, and calculate the flow split ratios as $\lambda_e = \tilde{\lambda}_e/\sum_{e'\in \hat{V}_{\text{out}}(q) } \tilde{\lambda}_{e'}$, $e\in \hat{V}_{\text{out}}(q)$, ensuring that $\sum_{e\in \hat{V}_{\text{out}}(q)}\lambda_e = 1$.

The flow split ratios provide insight into where the crowd flows after passing through a vertex. By normalizing with respect to incoming rather than outgoing edges, analogous ratios can be computed to describe where the crowd was before arriving at the vertex. With the weighted directed graph, $\hat{G}$, fully constructed, we next analyze the local flow Jacobian to extract semantic insights, offering a deeper understanding of the underlying dynamics and local features.

\subsection{Jacobian-Based Semantic Flow Analysis}
\label{subsec:local_linearization}
While the flow topology reconstruction of Sec.~\ref{subsec:flow_skeleton} effectively captures spatially structured flows, it can be complemented with additional analysis to reveal semantic information about the flow field, such as inflow (convergence), outflow (divergence), or rotational tendencies (clockwise/counterclockwise curl), even at points that are not vertices of $\hat{G}$. Moreover, for spatially diffuse flows, the extracted topology may lack meaningful structure. To this end, we next introduce an approach inspired by dynamical systems that identifies key regions without relying on topology reconstruction.

We can describe the dynamics of a particle at a location $\mathbf{x}=(x, y)$ in the flow field $\mathbf{\hat{v}}_{\text{unit}}(\mathbf{x}) = [\hat{v}_x(x, y), \hat{v}_y(x, y)]^T$ using a continuous-time equation $\frac{d\mathbf{x}}{dt} = \mathbf{\hat{v}}_{\text{unit}}(\mathbf{x})$. This governing equation can be expanded as a Taylor series~\cite{apostol1974mathematical} around a suitable point $\mathbf{x}_0 = (x_0, y_0)$ near $\mathbf{x}$, yielding $\frac{d\mathbf{x}}{dt} = \mathbf{\hat{v}}_{\text{unit}}(\mathbf{x}) \approx \mathbf{\hat{v}}_{\text{unit}}(\mathbf{x}_0) + \mathcal{J}(\mathbf{x}_0) \cdot (\mathbf{x} - \mathbf{x}_0)$, where
\begin{align} 
\mathcal{J}(\mathbf{x}_0) &= 
\begin{bmatrix} 
\frac{\partial \hat{v}_x}{\partial x} & \frac{\partial \hat{v}_x}{\partial y} \\
\frac{\partial \hat{v}_y}{\partial x} & \frac{\partial \hat{v}_y}{\partial y} 
\end{bmatrix}\bigg|_{(x_0, y_0)}. 
\end{align}

\begin{figure*}
    \centering
    \includegraphics[width=\linewidth]{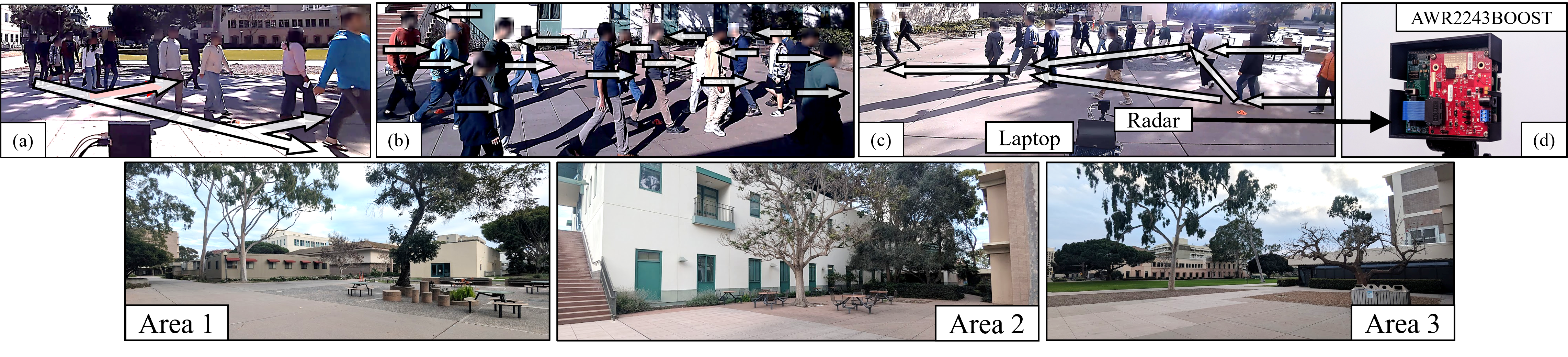}
    \caption{Experimental setup and environments: (Top) A mmWave MIMO radar transmits FMCW pulses, which reflect off individuals moving in (a, c) structured and (b) diffuse crowd flows (white arrows). Reflected signals are logged via Ethernet. (d) Front-facing view of the radar. (Bottom) Testing areas include open spaces with foliage (Areas 1 \& 3) and a courtyard with structural constraints (Area 2), where participants move freely without predefined paths or controlled pacing, mimicking real-world pedestrian dynamics. See color PDF for best viewing.}
    \label{fig:exp_setup}
\end{figure*}

The Jacobian, $\mathcal{J}(\mathbf{x}_0)$, describes how the flow varies around $\mathbf{x}_0$~\cite{perry1975critical, sanchez2012ordinary}, and it includes all information needed to calculate the 2D curl and divergence at $\mathbf{x}_0$. Specifically, we have~\cite{prieve2000course}
\begin{align}
\label{eq:curl_div}
\nabla\cdot\mathbf{\hat{v}}_{\text{unit}} = \frac{\partial \hat{v}_x}{\partial x} + \frac{\partial \hat{v}_y}{\partial y} \text{\ and\ }  
\nabla\times\mathbf{\hat{v}}_{\text{unit}} = \biggr(\frac{\partial \hat{v}_y}{\partial x} - \frac{\partial \hat{v}_x}{\partial y}\biggl),
\end{align}

\noindent for the divergence and curl, respectively. Intuitively, divergence quantifies the extent of gathering (negative) or dispersing (positive)
, while curl measures clockwise (negative) or counterclockwise (positive) rotation.

To characterize the local flow dynamics, we compute the curl and divergence scalar fields of the vector flow field. To do so, we must obtain the Jacobian, $\mathcal{J}(\mathbf{x})$, at each point, which in practice, is typically computed using finite difference methods. However, this approach is not robust to localized errors in the estimated flow and further suffers at the boundary of regions with zero flow (where $\mathbf{\hat{v}}_{\text{unit}}(\mathbf{x}) = \mathbf{0}$). We therefore propose a novel procedure for estimating $\mathcal{J}$ by solving a simple least-squares convex program. 

Specifically, recall that for a point, $\mathbf{y}$, in a neighborhood\footnote{If $\mathbf{\hat{v}}_{\text{unit}}(\mathbf{y}) = \mathbf{0}$, we do not include $\mathbf{y}$ in the neighborhood, as a flow of $\mathbf{0}$ indicates the point is not part of the region of interest.} $\mathcal{N}_L(\mathbf{x})$ around $\mathbf{x}$, we have $\mathbf{\hat{v}}_{\text{unit}}(\mathbf{y}) \approx \mathbf{\hat{v}}_{\text{unit}}(\mathbf{x}) + \mathcal{J}(\mathbf{x})\cdot(\mathbf{y} - \mathbf{x})$. The estimated Jacobian, $\mathcal{\hat{J}}(\mathbf{x})$, is then given  as 
\begin{equation}
    \mathcal{\hat{J}}(x), \mathbf{b}^*  \hspace{-0.05in} =  \hspace{-0.02in} \argmin_{J, \mathbf{b}}\hspace{-0.1in} \sum_{\mathbf{y}\in\mathcal{N}_L(\mathbf{x})} \hspace{-0.1in} \left\|\mathbf{\hat{v}}_{\text{unit}}(\mathbf{y})-( \mathbf{b} + J\cdot (\mathbf{y} - \mathbf{x}))\right\|^2,
\end{equation}
with $\mathbf{b}^*$ giving a smoothed estimate of $\mathbf{\hat{v}}_{\text{unit}}(\mathbf{x})$. We then use the estimated Jacobian, $\mathcal{\hat{J}}(\mathbf{x})$, to find the divergence and curl field at the point $\mathbf{x}$, as shown in Eq.~(\ref{eq:curl_div}). 

Having established this theoretical foundation for semantic analysis of flow fields, we next turn to experimental validation to demonstrate how our proposed crowd flow analysis methodology captures the essential characteristics of complex flow patterns in real-world scenarios. 

\begin{figure}
    \centering
    \includegraphics[width=\linewidth]{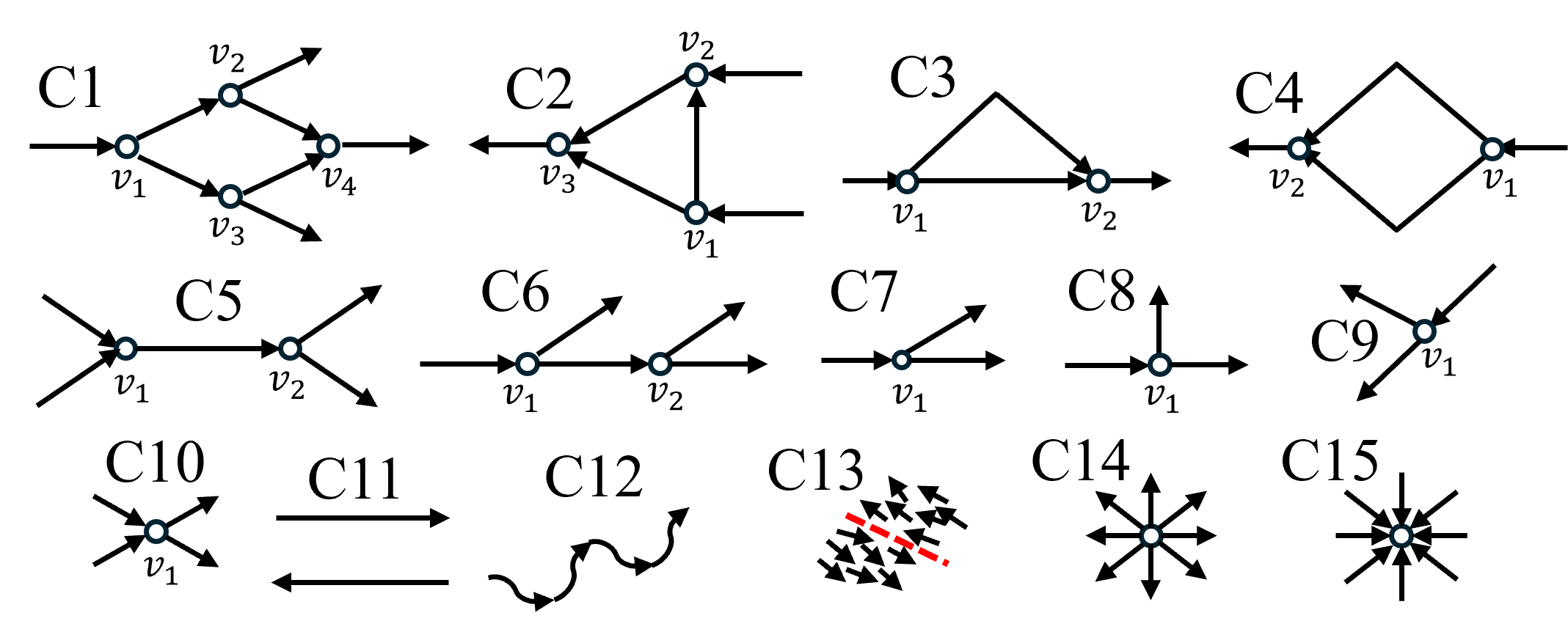}
    \caption{Graph representations of various crowd flow topologies tested using our proposed pipeline. Graphs C1 to C12 represent spatially structured flows, whereas C13 to C15 represent spatially diffuse flows. Arrows denote the flow paths and circles denote waypoints.}
    \label{fig:different_topologies}
\end{figure}

\section{Experimental Validation}
\label{sec:exp_val}
We next present the results of a comprehensive evaluation of our proposed system for crowd flow analytics through a series of real-world experiments. We first describe our experimental setup and testing areas before presenting empirical results from 21 experiments, with crowds of up to (and including) 20 individuals,\footnote{\textbf{This research has been reviewed and approved by our Institutional Review Board (IRB) committee.}} that validate our system's capability to reconstruct the crowd flow topology and further extract relevant spatial insights. Specifically, our approach in Sec.~\ref{sec:crowd_flow_analysis} recovers the flow topology in 17 of 18 (94\%) experiments focused on graph reconstruction, with all 18 experiments showing strong visual alignment with the ground truth. Even in the challenging case, our system successfully infers a majority of the flow structure (see Sec.~\ref{sec:discussion} for discussion). To quantify the accuracy of our reconstruction, we compute the one-sided Chamfer distance~\cite{liu2010fast} between the 2D embeddings of our estimated graph $\hat{G}$ and the ground-truth topology $G$, achieving an average distance of 0.45~m, well within the width of a typical pedestrian traffic lane~\cite{fruin1970designing}. Further, we achieve an edge orientation MAE of $8.8^\circ$ between the ground-truth and estimated edges, and a split ratio MAE of 0.1. Notably, these results are obtained using only a single mmWave radar board, and in environments with substantial foliage noise. We conclude by showing how the flow field curl and divergence can be used to localize regions that may correspond to boundaries between dominant flow directions or anomalous events within the sensing field of view.

\subsection{Experimental Setup}
\label{sec:exp_setup}
We assess our proposed system using a TI AWR2243BOOST off-the-shelf mmWave MIMO radar board~\cite{awr2243boostug}, as shown in Fig.~\ref{fig:exp_setup} (d). Operating at a base frequency of $f_0 = 76$~GHz, the radar transmits FMCW pulses with a $B = 5$~GHz bandwidth. To leverage its MIMO capabilities, we configure $N_{\text{TX}} = 3$ transmitters and $N_{\text{RX}} = 4$ receivers, constructing a $2 \times 8$ virtual array. However, given the limited elevation resolution (Sec.~\ref{sec:radar_pc_gen}), we collapse the vertical dimension, treating the array as a $1 \times8$ virtual uniform linear array. To improve azimuthal resolution, we apply zero-padding before performing angle-of-arrival estimation. The radar’s Channel State Information is recorded via a DCA1000EVM FPGA~\cite{dca1000evm}, which streams received chirp data to a laptop over Ethernet.  

\textbf{Experimental Areas:} We conduct experiments in three outdoor environments, illustrated in Fig.~\ref{fig:exp_setup} (Bottom). Area 1 is a 12 m $\times$ 12 m open space with seating on one side and a trailer on the other. Area 2 is a 21 m $\times$ 14 m semi-enclosed passage between two buildings, featuring outdoor seating, a large tree, and other structural elements that influence crowd movement. Area 3 is a 14 m $\times$ 11 m open area surrounded by substantial foliage. \textbf{While waypoints on the ground provide destination guidance, participants navigate freely between them without any prescribed paths or lanes, allowing for natural variations in movement while still adhering to the overall flow structure}. This fosters a dynamic and authentic crowd flow, capturing realistic movement patterns during data collection.

\begin{figure*}
    \centering
    \includegraphics[trim={0 0 0in 0}, clip, width=0.9\linewidth]{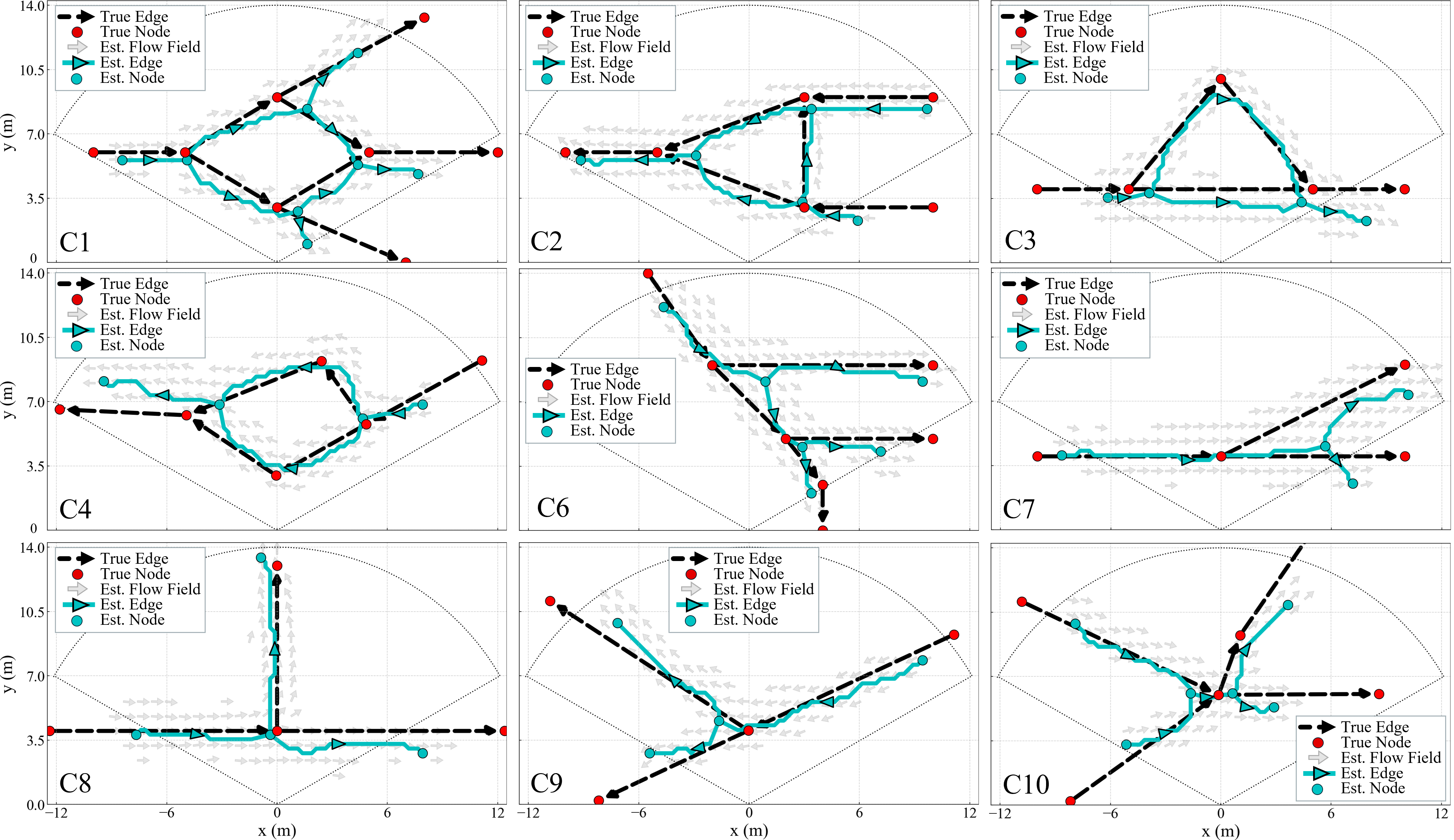}
    \caption{Sample estimated flow topology, $\hat{G}(\hat{V}, \hat{E})$, obtained from our proposed pipeline, compared with the true flow topology, $G(V, E)$. Our results capture the spatial distribution, connectivity patterns, and flow characteristics of the ground truth, and demonstrate the high-fidelity reconstruction of crowd flow topologies. The estimated flow field, $\mathbf{\hat{v}}_{\text{unit}}(\mathbf{x})$, has been downsampled for clarity of presentation. See color PDF for best viewing.}
    \label{fig:graph_reconstruction}
\end{figure*}

\subsection{Experimental Results}
\label{sec:exp_results}
We next present empirical results that validate the performance of our system across three experimental areas, accommodating crowd flows of up to (and including) 20 individuals, and with varying split ratios (see Sec.~\ref{subsec:flow_rate_split_estimation}). We model the observed area as a 30 m $\times$ 15 m rectangle, discretized at $0.25\ \text{m}$ intervals along each axis to align with the average biacromial width of an individual \cite{biacromialCDC}. Our implementation leverages OpenCV’s Lucas-Kanade algorithm (Sec.~\ref{subsec:pwf_generation})~\cite{opencv_lk}, with $\mathcal{N}_o(\mathbf{x})$ set as a $2\ \text{m} \times 2\ \text{m}$ square surrounding each point within the sensing field of view, based on maximum expected displacement between consecutive frames. For noise filtering, we set the Kolmogorov p-value (Sec.~\ref{subsec:ks_denoising}) threshold at $p_{\text{th}}= 0.15$. Additionally, we apply the following filtering parameters: minimum area of connected flow component (Sec.~\ref{subsec:flow_smoothing}): $\mu_S = 2\ \text{m}^2$, corresponding to the spatial footprint of small pedestrian groups~\cite{costa2010interpersonal}, and median filter window (Sec.~\ref{subsec:flow_smoothing}): $\mu_M = 0.25\ \text{m}^2$, matching the approximate area occupied by an individual \cite{fruin1987pedestrian}. For constructing $\hat{G}$ (Sec.~\ref{subsec:flow_skeleton}), we set $\Upsilon=1\ \text{m}$, and for split ratio calculation, we set the width of the edge bounding polygons to $\rho = 1\ \text{m}$, both aligned with standard lateral clearance maintained during pedestrian movement~\cite{gerin2005negotiation}. To compute the curl and divergence fields (Sec.~\ref{subsec:local_linearization}), we perform semantic flow analysis within a $|\mathcal{N}_L(\mathbf{x})| = \mu_M =\ 0.25\ \text{m}^2$ square around each point $\mathbf{x}$.

\textbf{Crowd Topology Reconstruction:} We conducted 18 spatially structured crowd flow experiments across three areas, with crowd sizes of up to (and including) 20 participants, and with a variety of flow split ratios. The ground-truth flow topology graphs, $G(V, E)$, for these experiments corre-
spond to configurations C1 through C12 in Fig.~\ref{fig:different_topologies}. Among these, topologies C4 and C9 are used in two additional experiments each, with distinct flow-splitting ratios to evaluate the sensitivity of our pipeline to varying crowd distribution patterns. C11 is used in two additional experiments where both flows move in the same direction, one with simultaneous and one with staggered starts, to test occlusion. These topologies were designed to emulate pedestrian movement in environments where structured flows naturally emerge due to underlying spatial constraints, while accommodating flow splitting and merging at points within the sensing field of view. In each experiment, the radar was located at the center of the bottom periphery of the area. We processed 30 seconds of radar data and applied our pipeline (Sec.~\ref{sec:radar_pc_gen},~\ref{sec:ff_synthesis} and~\ref{subsec:flow_skeleton}) to produce an estimate, $\hat{G}(\hat{V}, \hat{E})$ of the underlying flow topology, $G(V, E)$.

Our pipeline successfully reconstructed flow topologies, demonstrating strong visual alignment with the ground-truth topologies across 17 experiments. Fig.~\ref{fig:graph_reconstruction} illustrates samples of these high-fidelity reconstructions, which accurately capture the true crowd flow patterns, even for complex flow patterns, such as the branching structures in configurations C1, C2, and C6. However, our flow recovery fails in one unique case within C4. We discuss this scenario in more detail in Sec.~\ref{sec:discussion}, while for other C4 experiments, we successfully recover the flow topology (see Fig.~\ref{fig:graph_reconstruction}). 

\begin{table*}
\centering
\setlength{\tabcolsep}{3.1pt}
\begin{tabular}{|c|c|c|c|c|c|c|c|c|c|c|c|c|c|c|c|c|c|c|c|c|c|}
\hline
\textbf{Config.} & \multicolumn{2}{c|}{C10} & C9 & C9 & C9 & C8 & C7 & \multicolumn{2}{c|}{C6} & \multicolumn{2}{c|}{C5} & C4 & C4 & C3 & \multicolumn{3}{c|}{C2} & \multicolumn{4}{c|}{C1} \\
\hline
\hline
\textbf{Vertex} & $v_1^{\text{in}}$ & $v_1^{\text{out}}$ & $v_1$ & $v_1$ & $v_1$ & $v_1$ & $v_1$ & $v_1$ & $v_2$ & $v_1$ & $v_2$ & $v_1$ & $v_1$ & $v_1$ & $v_1$ & $v_2$ & $v_3$ & $v_1$ & $v_2$ & $v_3$ & $v_4$ \\
\hline
\textbf{True} & 0.50 & 0.50 & 0.20 & 0.50 & 0.80 & 0.25 & 0.25 & 0.33 & 0.50 & 0.53& 0.53 &  0.20 & 0.50 & 0.25 & 0.5 & 0.66 & 0.75 & 0.5 & 0.5 & 0.5 & 0.5 \\
\hline
\textbf{Est.} & 0.47 & 0.47  & 0.23 & 0.38 & 0.70 & 0.20 & 0.20 & 0.26 & 0.56 & 0.38 & 0.32 &  0.29 & 0.47 &  0.16 &  0.68 & 0.63 & 0.55 & 0.61 & 0.69 & 0.51 & 0.33\\
\hline
\end{tabular}
\caption{Estimated flow split ratios at vertices of flow bifurcation across configurations C1–C10. For each vertex, we report the ratio for one of the edges, $\lambda_e$, with the ratio for the other edge given by $1-\lambda_e$. Our method achieves a Mean Absolute Error (MAE) of 0.1, highlighting its efficacy in capturing details on the flow dynamics.}
\label{tab:flow_rate_split_table}
\end{table*}

To further assess our proposed approach, we evaluate the accuracy of the graph's embedding in 2D space. Determining a suitable metric for this task requires careful consideration, as participants move freely between nodes to maintain realistic crowd behavior, rather than following strict straight-line paths defined by the flow topologies. We therefore use a metric that is robust to small local shifts, while still penalizing large structural discrepancies. To this end, we evaluate the one-sided Chamfer distance $d_{\text{avg}}$~\cite{chen2022unist, liu2010fast}, given by the average minimum distance between a point on the estimated graph, $\hat{G}$, and any point on the true graph, ${G}$. More specifically, let ${\mathcal{P}}$ and $\hat{\mathcal{P}}$ denote the set of points that constitute the 2D embeddings of $G$ and $\hat{G}$, respectively. We then calculate $d_{\text{avg}} = (1/|\hat{\mathcal{P}}|) \int_{\hat{\mathcal{P}}} \min_{p\in\mathcal{P}} ||\hat{p} - p||_2\,d\hat{p}$, where $|\hat{\mathcal{P}}|$ is the sum of the edge lengths of $\hat{G}$. We calculate $d_{\text{avg}}$ for each of the 17 successfully reconstructed graphs, obtaining a mean $d_{\text{avg}}$ of 0.45~m. We emphasize that although the ground-truth edge is represented as a 1D curve, in reality, the width of a typical pedestrian traffic lane is 0.76~m~\cite{fruin1970designing}. Thus, our estimated edges remain well within this
natural width.

To further quantify accuracy, we associate edges in $G$ and $\hat{G}$, and characterize errors in flow direction estimation by calculating the angular orientation error between the edges. For the 62 edges present in the successfully reconstructed graphs, we achieve an edge orientation MAE of $8.8^\circ$. Overall, these results demonstrate the robustness of our approach both in reconstructing the abstract flow graph and reliably localizing its embedding in 2D space.

\textbf{Flow Split Ratio Estimation:}
We next demonstrate an application of this extracted directed geometric graph, $\hat{G}$, to develop an understanding of the relationships between the regions of the observed space. Specifically, we estimate the flow split ratios into (out of) a node in $\hat{G}$ along each of the incoming (outgoing) edges using our proposed pipeline from Sec.~\ref{subsec:flow_rate_split_estimation}. In each experiment that featured a splitting or merging point (configurations C1 through C10 of Fig.~\ref{fig:different_topologies}), we set flow split ratios beforehand and assigned participants to specific flow edges accordingly to ensure that the real crowd flows reflected these ratios. We note that the participants were not pre-grouped by their assigned directions so that they remained naturally mixed until they reached the splitting point. Table~\ref{tab:flow_rate_split_table} compares our estimated flow split ratios to the true values, highlighting a great performance, with a Mean Absolute Error (MAE) of 0.10 across all experiments. Furthermore, when considering the simpler branching topologies in configurations C7-C9, the MAE drops to 0.05. For configurations C9 and C4, we conducted multiple experiments with different split ratios. Our estimation method not only accurately captured these variations but also effectively distinguished relative changes in the flow split ratio within the same topology. Importantly, flow split ratio estimation relies on an accurate $\hat{G}$, as its topology is essential for linking inflow with outflow directions.

\begin{figure*}
    \centering
    \includegraphics[width=\linewidth]{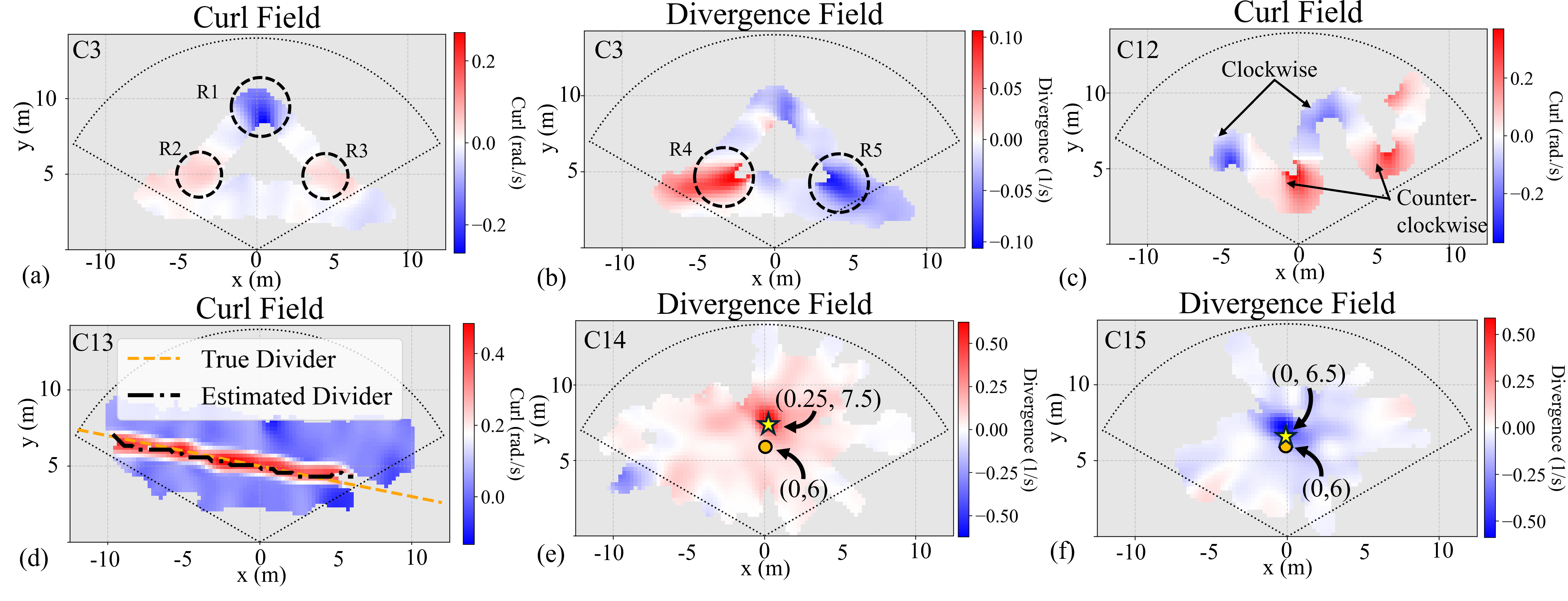}
    \caption{Sample curl and divergence fields for structured (top) and diffuse (bottom) crowd flows. (a) High negative curl in R1 indicates an abrupt clockwise turn. (b) High positive (negative) divergence at R4 (R5) indicates a split (merge) in the flow. (c) Distinct curl regions correspond to clockwise and counterclockwise motion. (d) A narrow region of strong curl corresponds to the line dividing the two flow directions (orange: true, black: estimated). (e) Strong positive divergence (yellow star) aligns with the true divergence point (orange circle). (f) Strong negative divergence (yellow star) aligns with the true dispersion point (orange circle). See color PDF for best viewing.}
    \label{fig:curl_div_heatmaps}
\end{figure*}

\subsection{Semantic Insights from Flow Fields}
\label{subsec:semantic_exp}
In results above, we have shown how to use the estimated flow field, $\hat{\mathbf{v}}_{\text{unit}}(\mathbf{x})$, to reconstruct the flow topology. Com-
plementary to this reconstruction process, we next demonstrate the effectiveness of our semantic flow analysis pipeline (Sec.~\ref{subsec:local_linearization}) in methodically discovering regions in the flow that exhibit significant local divergence or curl properties. These properties are important as they correspond to key crowd behaviors: high positive (negative) divergence is associated with flow splitting (merging), while high positive (negative) curl is associated with acute counterclockwise (clockwise) turns in the flow. We note that an additional low-pass filter is applied to $\mathbf{\hat{v}}_{\text{unit}}(\mathbf{x})$, to enhance the visualization of curl and divergence fields. We first present this analysis in the context of structured crowds before showing that this approach generalizes to diffuse crowds as well. 

\textbf{Identifying Key Areas in Spatially Structured Crowd Flows:}
We next demonstrate how divergence and curl fields provide semantic information in structured crowd flows. To illustrate, Fig.~\ref{fig:curl_div_heatmaps} (Top) shows curl and divergence fields for configurations C3 and C12 in Fig. \ref{fig:different_topologies}.  In Fig.~\ref{fig:curl_div_heatmaps} (a), the curl field for C3 shows intense negative curl in region R1, where the crowd flow makes an abrupt clockwise (CW) turn, with less intense positive curl around R2 and R3 indicating less acute counterclockwise (CCW) motion. Fig.~\ref{fig:curl_div_heatmaps} (b) shows the divergence field for C3, which indicates strong positive divergence at R4, where the flow splits, and strong negative divergence at R5, where the flow merges. Finally, Fig.~\ref{fig:curl_div_heatmaps} (c) shows the curl field for configuration C12, with negative and positive curl naturally segmenting the flow into regions of CW and CCW movement. Thus, our pipeline provides semantic understanding of the flows to complement the reconstruction of the flow topology discussed earlier, and we next illustrate how this analysis generalizes to diffuse flows.

\textbf{Detecting Regions of Interest in Spatially Diffuse Crowd Flows: }
The semantic flow analysis pipeline (Sec.~\ref{subsec:local_linearization}) also provides useful insights into the structure of more diffuse flows. For such crowds, the KS denoising step of Sec.~\ref{subsec:ks_denoising} requires parameter adjustment given the more random nature of the true flows. We therefore set the threshold parameter, $p_{\text{th}} = 1$ for these scenarios, which allows us to recover estimated flow fields rich enough to provide structural insights into crowd behavior. One can easily determine when to use each parameter, as we discuss in Sec.~\ref{sec:discussion}.

Fig.~\ref{fig:curl_div_heatmaps} (Bottom) presents the curl and divergence fields for configurations C13, C14, and C15 in Fig.~\ref{fig:different_topologies}, respectively. In C13, participants move in two diffuse lane-like flows separated by a boundary (red dotted line in Fig.~\ref{fig:different_topologies}, C13, see also Fig.~\ref{fig:exp_setup} (b)), where the flow direction undergoes an abrupt 180-degree reversal. In C14, upon receiving a signal, participants evacuate radially outward from a designated starting position, simulating a crowd panic scenario. Conversely, C15 captures participants converging toward a central destina-
tion from multiple directions, modeling a gathering scenario. As observed in Fig.~\ref{fig:curl_div_heatmaps} (d), the curl field of C13 distinctly highlights the flow separator where the two lane-like streams un-
dergo direction reversal. Similarly, the location of the highest intensity peak in the divergence field of C14 lies close to the source of the crowd panic event, aiding in localizing anomaly origins in real-world scenarios (Fig.~\ref{fig:curl_div_heatmaps} (e)). In C15, a region of strong negative divergence corresponds to the collective gathering location (Fig.~\ref{fig:curl_div_heatmaps} (f)). Thus, our estimated flow fields provide sufficient detail to recover structural information via semantic analysis, even for diffuse flows.

\section{Discussion and Future Work}\label{sec:discussion}
We next benchmark our approach against relevant methods and discuss directions for future work.

\begin{figure}
    \centering
    \includegraphics[trim={0.2in, 0, 0, 0}, clip, width=\linewidth]{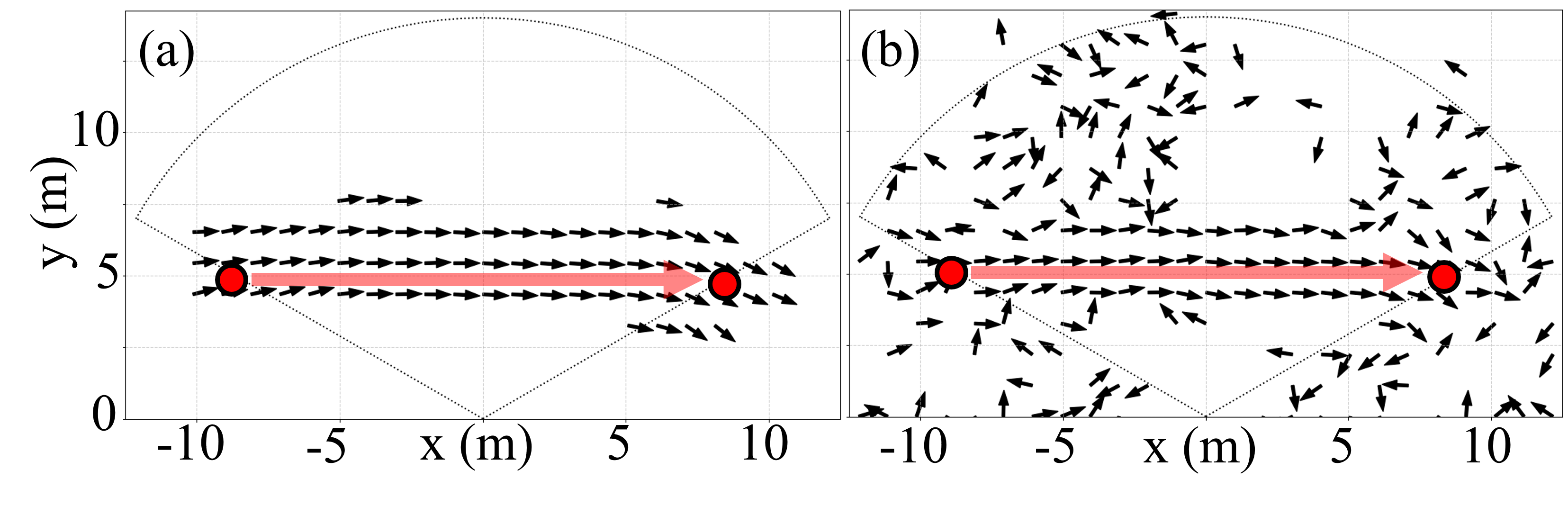}
    \caption{Motivating the need for new flow field modeling for mmWave crowd sensing: (a) our proposed flow field model of Sec.~\ref{sec:ff_synthesis} results in significantly better performance as compared to (b) the well-known Lucas-Kanade optical flow algorithm in vision. Both fields are downsampled for visibility.}
    \label{fig:denoising_perf}
\end{figure}

\noindent\textbf{Comparison with Prior Work:}
Since there is no prior work on mmWave-based crowd flow analytics, as discussed earlier, we next compare specific components of the pipeline to the related work in literature. For instance, direct application of classical optical flow methods from the area of vision produces visibly noisy flows, as shown in Fig.~\ref{fig:denoising_perf}. Furthermore, mmWave flow extraction pipelines developed in other domains, e.g.,~\cite{Ding2024milliFlow}, do not generalize for crowd flows, and processing our point clouds through~\cite{Ding2024milliFlow}, for instance, produced unusable flow fields due to the scale disparity between single human activity and crowd motion. These comparisons highlight our proposed flow field analysis as a significant advancement in the emerging domain of mmWave crowd flow analytics.

\begin{figure}
    \centering
    \includegraphics[width=0.7\linewidth]{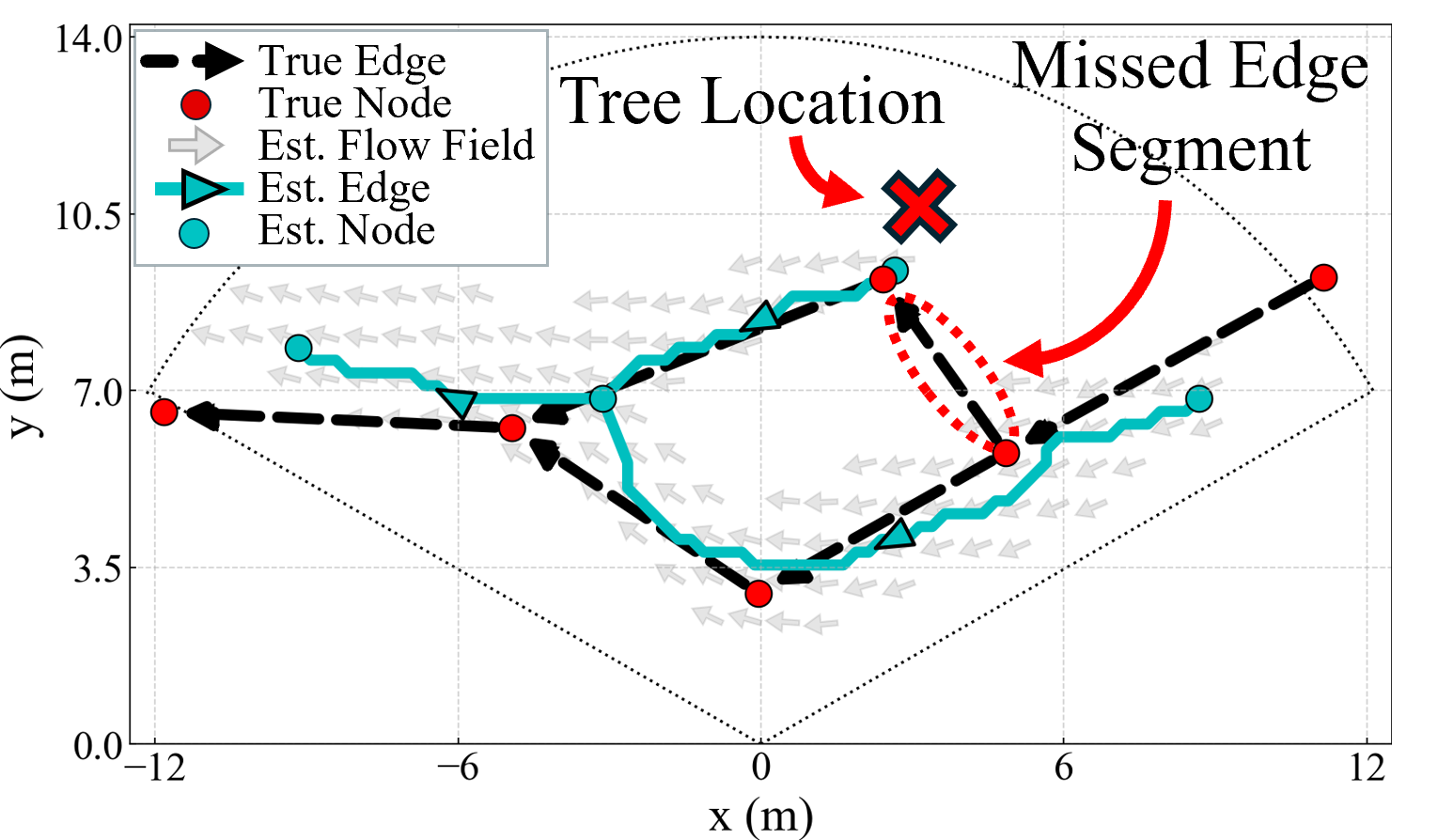}
    \caption{Case of incomplete graph reconstruction on C4. Despite significant noise and sparse flows on the upper route, nearly all of the flow topology is recovered, except for one edge.}
    \label{fig:bad_recon}
\end{figure}

\noindent\textbf{The Case of Unsuccessful Flow Topology Extraction:}
As stated in Sec.~\ref{sec:exp_results}, our pipeline did not completely recover the flow topology in one of the 18 structured-crowd experiments. Specifically, we ran 3 experiments on configuration C4 with 80\%, 50\%, and 20\% of the crowd taking the upper route at $v_1$, and in the last of these, flow topology recovery was partially successful (missed one edge), as shown in Fig.~\ref{fig:bad_recon}. Two key factors caused this incomplete recovery, both stemming from the highly unequal split ratio. First, significant occlusion, caused by the majority of the crowd traveling along the lower route, reduced visibility of the upper route. Second, a tree along the upper route produced persistent noisy returns that dominated the sparse crowd-related signals. Consequently, points pertaining to the true crowd flow were discarded as noise. For such cases, a more granular analysis over smaller temporal windows is an avenue for future work.

\noindent\textbf{Distinguishing between Structured and Diffuse Crowds:} Our proposed graph construction framework can be used to differentiate between structured and diffuse crowds, as graphs constructed for diffuse crowd flows exhibit distinguishing characteristics. One such indicator is the inconsistency between the directions of the inferred graph edges and those of the underlying estimated flow field, as measured by, \textit{e.g.,} the average cosine similarity. Structured crowds show much higher consistency, as expected. A comprehensive exploration of such distinctive factors is a future work direction.   

\section{Conclusion}
This work represents a significant advancement in crowd flow analytics using commodity mmWave MIMO radar. Our method successfully bridges the gap between raw radar data and meaningful crowd behavior analysis by transforming radar point clouds into directed geometric graphs through interpretable flow fields. The high accuracy achieved across three areas validates the robustness of our approach, with high-fidelity graph reconstruction of the underlying flow structure in considerably complex crowd patterns, and accurate flow split ratio estimation with an MAE of 0.1.
Key contributions of our work include a novel framework for the high-fidelity generation of mmWave crowd flow fields, a new approach for extracting the underlying crowd topology, and further quantifying flow ratios, as well as the application of vector field analysis to extract meaningful semantic information from crowd flows. By analyzing the Jacobian of the flow field, particularly through the divergence and curl, we provide deeper insights into crowd behavior, applicable to both spatially structured and diffused crowds. This is particularly valuable for characterizing spatially diffuse flows where topology reconstruction alone proves insufficient. Our extensive experimental validation confirms the effectiveness of our methodology, establishing a solid foundation for future advancements in mmWave-radar-based crowd flow analysis.

\section{Acknowledgements}
This work is funded in part by ONR award N00014-23-1-2715 and in part by NSF CNS award 2215646.

\bibliographystyle{ACM-Reference-Format}
\bibliography{main}

\end{document}